# Volumic deformation of human cornea under pressure

Chloé Giraudet[a,b], Qian Wu[a,b], Jean-Marc Allain[a,b,#]

[a] LMS, CNRS, Ecole Polytechnique, Institut Polytechnique de Paris, Palaiseau, France
[b] Inria, Palaiseau, France

# Corresponding author. Email: jean-marc.allain@polytechnique.edu

**Abstract**

The cornea, as the outer element of the human eye, plays a pivotal role in vision. Any defects in its shape can result in visual impairments. Mechanical defect manifest as shape defects, as the cornea is under to pressure. Our study presents the first comprehensive observation of human corneal deformation throughout its entire thickness during an inflation test, through Optical Coherence Tomography (OCT). Horizontal deformation reveals depth-dependent heterogeneity, suggesting that the cornea's posterior part is softer than the anterior part. Vertical deformation was observed at levels significantly higher than expected and exhibited depth-dependent heterogeneity, delineating three distinct regions. The central region of the cornea initially experienced rapid swelling, possibly due to osmotic effects, followed by increasing compressions as pressure rose. Conversely, the anterior and posterior regions showed no sign of swelling, and a difference in the compressive response, possibly due to difference in stiffness. Our study shows the complexity of human corneal mechanics, highlighting strong anisotropy and depth-dependent behavior, and implicating osmotic and poroelastic properties.

## 1. Introduction

The cornea is the tissue that forms the outer part of the eye and plays an essential role in vision, both through its transparency and its shape, which accounts for 2/3 of the eye's refractive power. This shape is ensured both by natural curvature and by the cornea's biomechanical properties, as the cornea is subjected to intraocular pressure from the aqueous humor. Surgeries or pathologies may change the cornea mechanical properties, with an impact on its shape [1,2].

The cornea has already been extensively studied from a mechanical point of view, utilizing most of the techniques available to mechanics: uniaxial or biaxial tractions [3–6], inflation tests [7–11], indentation [12,13], and other more complex techniques [14,15]. These experiments have shown a non-linear stress-strain behavior, similar to other collagen-rich tissues. Indeed, the cornea is mainly formed by the stroma, which accounts for 90% of its thickness, and is made up of a stack of collagen fibril lamellae, reminiscent of plywood 4/25/2024 10:32:00 AM. This structure is heterogeneous laterally and in thickness [16–18], leading to questions about the heterogeneity of the tissue's mechanical properties with depth. Tensile or indentation experiments have shown that tissue stiffness decreases with depth from the outer surface [5]. However, static elastography observations show a difference between the central zone, and the anterior and posterior zones [19].

The difficulty of corneal testing comes from the complexity of the tissue, which combines a heterogeneous material with a naturally curved structure and variable thickness. Inflation tests are more physiological, even though strip tensile tests yield similar behavior. This is why we have chosen to perform inflation tests. Usually, inflation tests track the movement of the corneal apex (i.e., its highest point) as a function of the pressure applied in the posterior chamber [7]. Notably, these are the only inflation measurements conducted on human corneas. There are also stereo correlation measurements on bovine corneas [8] and elastography measurements on rat corneas [19].

In this paper, we propose a full volume deformation of human corneas during an inflation test. To achieve this, we combined a conventional inflation device with imaging using an Optical Coherence Tomograph (OCT). Utilizing natural contrast, we quantified deformation at different pressure levels using Digital Volume Correlation. Our observations reveal significantly different behavior in directions tangent to corneal surfaces compared to the thickness direction. The tangential behavior indicates that the inner part of the cornea deforms more than the outer part. The thickness direction appears to be mainly controlled by water exchange, driven by either osmotic or mechanical pressure, demonstrating the poro-elastic behavior of the tissue, governed by its microstructure.

## 2. Materials and Methods

### 2.1. Sample preparation

This study was conducted in accordance with the principles outlined in the Declaration of Helsinki and adhered to international ethical standards for the use of human tissues. Handling of human corneas was duly reported to the French regulatory authority (CODECOH agreement DC-2019-3638). Seven human corneas were acquired from the French Eye Bank (Banque Française des Yeux, BFY), which were deemed unsuitable for transplantation and granted authorization for scientific research by the donor families. These corneas were preserved in Stemalpha 2 culture medium (StemAlpha, France) and maintained in an incubator set at 31°C. Subsequently, they were immersed in Stemalpha 3 medium (StemAlpha, France) for 48 hours to induce deswelling. Table 1 presents the characteristics of the corneas including age, sex, thickness at the first step (at the apex), and endothelial density.

*2.2. Inflation tests*

The cornea was then secured by its sclera onto an anterior chamber (Artificial single-use chamber, Moria, France), which was connected to a homemade injector designed to deliver a known volume at a controlled rate (see Figure 1).

A pressure sensor (ATM.1ST, STS, France - range: 100kPa, accuracy 0.1Pa) was positioned upstream of the anterior chamber to record pressure readings every second. Special attention was given to positioning the sensor at the level of the cornea to prevent any hydrostatic pressure differential associated with the water column; thus, the measured pressure accurately reflected the pressure within the chamber.

Stemalpha 3 was injected to minimize changes in the osmotic environment. The fluid was circulated within the chamber until all air bubbles were expelled, following which the flow was halted, and the outlet of the chamber was sealed. Once the chamber was sealed, the injection proceeded at a rate of 0.01 cm$^3$/s. Pressure was increased incrementally in steps of 1.3-2 kPa (10-15 mmHg) up to 20 kPa (150 mmHg), then decreased back to physiological levels (2-2.7 kPa, 15-20 mmHg), before being increased again to 21 kPa (160 mmHg). Imaging was conducted at each pressure step, with the injection paused during imaging.

*2.3. OCT imaging*

The cornea was imaged using an Optical Coherence Tomograph (SD-OCT Ganymede – SP 5, Thorlabs, Germany) (see Figure 1). To ensure proper hydration of the anterior surface of the cornea, a drop of ophthalmic gel (Lacrigel 0.2%, Europhta Laboratory, Monaco) was placed between the OCT and the cornea. At each pressure step, both 2D and 3D images were acquired. While capturing a 2D image was nearly instantaneous, obtaining 3D images, which are stacks of hundreds of 2D images, took up to 2 minutes.

The 2D images were acquired along the Superior-Inferior (SI) and Naso-Temporal (NT) axis of the cornea, with a field of view of 8 mm (1333 px) in length and 1.5 mm (1024 px) in depth (see Figure 2). The 3D images were acquired around the apex, with a volume of 2.5x2.5x1.3mm (416x416x885 vx), the last dimension representing the depth (see Figure 2). The depth provided here is adjusted by the acquisition software to accommodate the refractive index of the medium: the OCT measures optical distance rather than real length. We selected n=1.33, the refractive index of water (for the ophthalmic gel). The cornea has a refractive index closer to 1.37, resulting in a 3% overestimation of the depths.

### *2.4. Strain measurement*

These images were then analyzed using digital volume/image correlation (DIC / DVC) software, CMV and CMV3D [20,21]. CMV and CMV3D are local correlation software: 2D/3D correlation subsets are taken at 30x30 $px^2$ and 30x30x30 vx, respectively, with search domains of 15x15 $px^2$ and 15x15x15 vx, respectively.

For the 2D images, the axes used for correlation were: (i) the horizontal axis x corresponding to either SI or NT direction, and (ii) the vertical axis z representing the depth direction of the cornea. For the 3D images, the axes used for correlation were: (i) the in-plane axes x and y corresponding respectively to the SI and NT directions, and (ii) the vertical axis z representing the depth direction of the cornea.

The correlation provided the position of each subset at different pressure steps. From these positions, the Green-Lagrangian strain tensor **e** was computed.

### 3. Results

We utilized our OCT images to quantify volume deformation. Digital Volume Correlation (DVC) provides access to the full strain tensor, whereas Digital Image Correlation (DIC) provides access to only 2D tensors but over a larger field of view.

Firstly, we analyzed horizontal deformations. Near the apex, this approximates tangent deformation. Figure 3 depicts horizontal strain maps at different depths (1/6, 1/3, 1/2, 2/3, and 5/6 of the thickness) for one representative cornea at the end of the first inflation. At each depth level, we did not observe significant heterogeneity, except in the posterior region where some lines of higher deformations were observed. Thus, it seems appropriate to use mean values. Figure 4a illustrates the evolution of mean horizontal strains with depth on this cornea, along with their standard deviations, computed from the volume deformation at the end of the first inflation. We observed progressive deformation increase from the anterior surface to the posterior surface. The standard deviation appeared to abruptly increase in the deeper half of the cornea, reflecting the emergence of lines of higher deformation. Shear remained near zero at all depths, as expected for biaxial loading, with increased variability in the deeper half.

Figure 4b displays the evolution of mean horizontal deformations at different pressure steps for all depths. We observed that deformations followed the pressure, increasing, decreasing, and then increasing again at all depths.

The same analysis was performed on the 2D images, where only one component of the strain tensor ($e_{xx}$ or $e_{yy}$) was accessible, albeit over a larger field of view. Figure 5a illustrates the horizontal strain maps for the two 2D images at the end of the first inflation. Similar observations to the volume analysis were made in the middle of the cornea. However, heterogeneity of the pattern near the cornea's border was noted, exhibiting a symmetric effect with the right border under compression and the left border under tension.

Next, we examined vertical deformations. Figure 6 presents maps of vertical strains at different depths (1/6, 1/3, 1/2, 2/3, and 5/6 of the thickness) for the same cornea at the end of the first inflation. For shear, patterns similar to horizontal ones were observed. However, the $e_{zz}$ component displayed high compression in the upper part of the cornea (down to -20%), which dissipated with depth, exhibiting a roughly homogeneous pattern.

Figure 7a illustrates the evolution of vertical strains with depth on this cornea, along with its standard deviations at the end of the first inflation. Similar to the horizontal component, shear displayed a mean value around 0, with increased standard deviation with depth.

For the vertical component, three distinct regions were observed. Starting from the anterior surface, strain became increasingly negative (the tissue more compressed) up to around 1/3 of the depth. Subsequently, strain continued to increase, indicating decreasing compression, until it became slightly stretched. Then, in the deeper region (near 2/3 of the depth), strain decreased again, becoming slightly compressed once more.

Figure 7b demonstrates the evolution of mean vertical strains for different pressure steps at all depths. Shear was small and primarily followed pressure. However, for vertical strain, the tissue was under significant stretch, particularly around 1/3 of the depth. Subsequently, the tissue became progressively more compressed from smaller depths towards deeper regions. At all pressure levels, the anterior surface appeared less compressed than the tissue just behind it, while the posterior surface appeared less stretched than the tissue just above.

The same analysis was conducted on the 2D images. Figure 5 illustrates vertical strain maps for the two 2D images at the conclusion of the first inflation. Similar observations to the volume analysis were made in the middle of the cornea. However, compression seemed to diminish as we reached the borders. Shear levels significantly increased at the border, exhibiting a symmetric effect similar to horizontal strain.

4. Discussion

We present the first observations of human corneal deformation throughout its full thickness during an inflation test. To achieve this, we utilized the intrinsic signal of the cornea imaged with an Optical Coherence Tomograph (OCT). While numerous past publications have explored the mechanical properties of the cornea, both human and animal [5,7,8,19], they have typically focused on either tangential or transverse deformations separately. By employing OCT, we were able to simultaneously observe the complete deformation in volume.

In the central region of the cornea, we observed homogeneous tissue deformation, consistent with the tissue's homogeneous microstructure [16]. However, this homogeneity was less evident in the posterior third, where significant heterogeneity emerged (see Figure 4). Particularly, we observed high deformation lines in the horizontal strains (see Figure 3). These lines may be attributed to the fact that our corneas were slightly thicker than the physiological thickness, resulting in folds on the posterior surface (see Figures 2 and 5).

The overall horizontal deformation at the end of the first inflation ramp was around 4%, falling within the range of a few percent expected from a homogeneous spherical cap. We noted that horizontal strains exhibited heterogeneity with depth, progressively increasing (see Figure 4). This contradicts what is expected for a homogeneous spherical cap under pressure and suggests a decrease in tangential (or horizontal) stiffness of the cornea with depth, consistent with findings in the literature [5].

Surprisingly, we observed significant deformation in the vertical direction, surpassing the expected levels, and in particular the incompressibility assumption (see Figures 6 and 7), as it reached up to 20% of compression with respect to 4% in the other directions.

This strong compression explained our observations of shear and horizontal strains at the borders of 2D images (see Figure 5). At the cornea's edge, the direction of corneal thickness no longer aligned with the image base's measurement directions. Closest to the edge, the angle between the cornea and the vertical was 27°. Considering a dilation tangent to the surface of around 4% and a thickness contraction of 20%, factoring in the angle on the measurement base led to an estimated vertical deformation of around 13% and a horizontal deformation of around 4%, consistent with our observations (see Figure 5).

This compression exhibited strong depth heterogeneity, with three regions, each covering approximately 1/3 of the cornea (see figure 7). Initially, the central region experienced swelling, progressively diminishing as pressure increased. This is in agreement with other observations in the literature [19]. This swelling disappears when the pressure is increased and didn't reappear upon pressure reduction. We assume that this swelling arises from an osmotic effect; the cornea was immersed in Stemalpha 3 medium for 2 days, osmotically distinct from the lacrigel atop the cornea. Thus, we consider the swelling to result from rapid water flux from the lacrigel to the stroma, facilitated by the ex-vivo cornea's poor epithelial quality.

As pressure on the cornea increased, the applied force progressively displaced the inner fluid, leading to significant compression of the middle part. This compression cannot be solely attributed to osmosis as we didn't change the surrounding medium. As the cornea volume strongly decreased, we suggest that this compression predominantly represents a poroelastic effect. Mechanical loading on the tissue reduces pores and drives outward water flux, resulting in substantial volume reduction.

We observed that in the anterior region, strain progressively decreased from the surface to one-third of the depth (see Figure 7). So, its gradient of strain with depth is opposite to the ones in the middle region. This feature aligns with our poroelastic interpretation if the anterior part is stiffer than the central one. This stiffness difference may come from a more out-of-plane organization of the lamellae near the anterior surface [17,18].

In the posterior region, strain progressively increased from the surface to two-thirds of the depth (see Figure 7). So, its gradient of strain with depth is similar to the anterior region, and opposite to the central region. This trend is challenging to explain based solely on microstructural observations. It may be related to endothelial water control, which regulates flux on the lower border, although we lack evidence to support this hypothesis. Alternatively, it may arise from different mechanical properties, but there is no literature evidence supporting a stiffer deeper region.

5. Conclusion

Our observations of the human cornea's volume deformation reveals that its ex-vivo mechanical properties are complex. They exhibit strong anisotropy, displaying different behaviors in the vertical direction compared to the tangential direction. The tangential direction appears significantly stiffer and almost elastic. We note a gradient of tangential deformation with depth, which can be attributed to a softer tissue in the posterior region than on the anterior region. Simultaneously, we observe pronounced deformation in the vertical direction, characterized by heterogeneous responses, showing three distinct regions than are not observed in the microstructure. The vertical deformation is likely to be controlled by a mix of osmotic effects, and poroelastic behavior.

**Fundings**

This work was supported by Agence Nationale de la Recherche (ANR-21-CE19-0010-01) and Ecole Polytechnique (interdisciplinary project METIS.)

**Declaration of Competing Interest**

The authors declare that they have no known competing financial interests or personal relationships that could have appeared to influence the work reported in this paper.

**Acknowledgments**

The authors thank the Banque Française des Yeux (BFY, Paris) for providing human corneas, Vincent de Greef for his help in maintaining the inflation set-up, Hakim Gharbi, Simon Hallais, and Martin Genet for their help in the image correlation analysis, Clothilde Raoux, Gaël Latour, and Marie-Claire Schanne-Klein from LOB (IP Paris), and Juliette Knoeri and Vincent Borderie, from CHNO 15-20 (Inserm, Sorbonne University) for fruitful discussions.

Figures and Table captions:

**Table 1:** Characteristic of the cornea used in the study.

**Figure 1:** Experimental set-up. Left: phography of the cornea on the injection chamber under the OCT. Right, scheme of all the different elements of the set-up

**Figure 2:** Imaging through by the OCT. Upper part: scheme of the locations of the planes (in red/blue) and of the volume (orange) with respect of the cornea. Lower part: examples of obtained images. In red, the SI cross-section, and its location on the cornea on the pressure chamber, viewed from the top. In blue, the NT cross-section, and its location on the cornea on the pressure chamber, viewed from the top. In orange, a cross-section of the volume, and its location on the cornea on the pressure chamber, viewed from the top.

**Figure 3:** Horizontal strain maps at different depths (1/6, 1/3, 1/2, 2/3, and 5/6 of the thickness) for Cornea 24167, at the end of the first inflation

**Figure 4:** Evolution of the mean horizonal strains on the Cornea 24167. (A) Evolution of the strains with depth at the end of the first inflation. The confidence interval of the data is given by mean ± standard deviation. (B) Evolution of the mean strains with pressure steps and depths.

**Figure 5:** Strain maps from 2D images, at the end of the first inflation. From the top to the bottom: (A) horizontal strain in SI cross-section (x direction), and in the NT direction (y direction), (B) vertical strain in SI cross-section (x direction), and in the NT direction (y direction), (C) shear strains in SI cross-section (x direction), and in the NT direction (y direction), and (D) pressure and volume versus time. The vertical line indicates when the images have been taken.

**Figure 6:** Vertical ($e_{zz}$) and shear ($e_{xz}$ and $e_{yz}$) strain maps at different depths (1/6, 1/3, 1/2, 2/3, and 5/6 of the thickness) for Cornea 24167, at the end of the first inflation.

**Figure 7:** Evolution of the mean vertical ($e_{zz}$) and shear ($e_{xz}$ and $e_{yz}$) strains on the Cornea 24167. (A) Evolution of the strains with depth. The confidence interval of the data is given by mean ± standard deviation. (B) Evolution of the strains with pressure steps and depths. The pressure steps are showed at the bottom.

**Table 1:**

| Cornea Id | Age | Genre | Endothelial cell density (cell/mm²) | Thickness at Apex (mm) |
|---|---|---|---|---|
| 23769 | 72 | F | 1600 | 0.67 |
| 23809 | 81 | F | 2650 | 0.69 |
| 23831 | 66 | M | 2750 | 0.75 |
| 23849 | 82 | M | 1950 | 0.67 |
| 24109 | 87 | F | 2400 | 0.91 |
| 24149 | 74 | M | 1900 | 0.72 |
| 24167 | 72 | M | Impossible to count | 0.71 |

**Figure 1:**

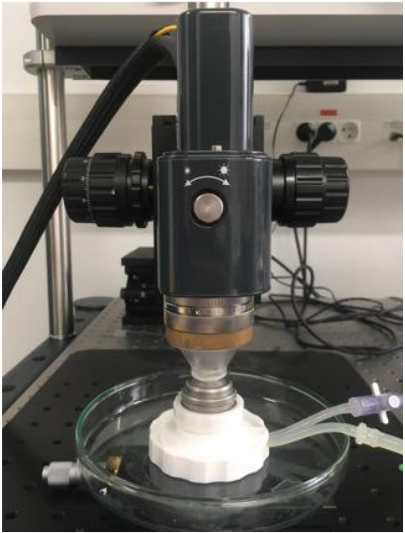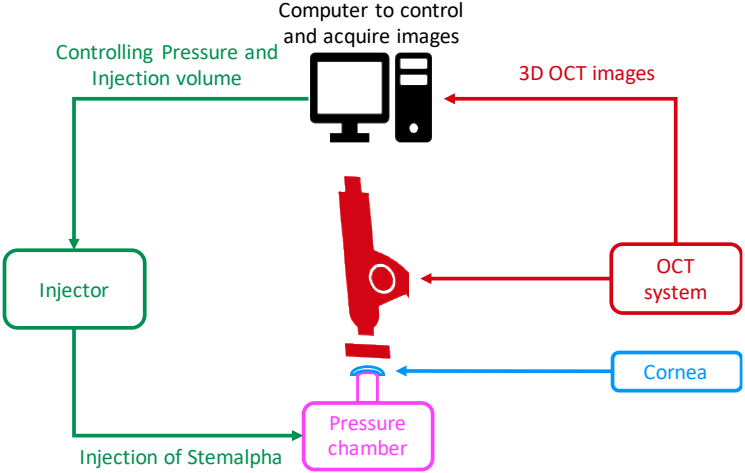

**Figure 2:**

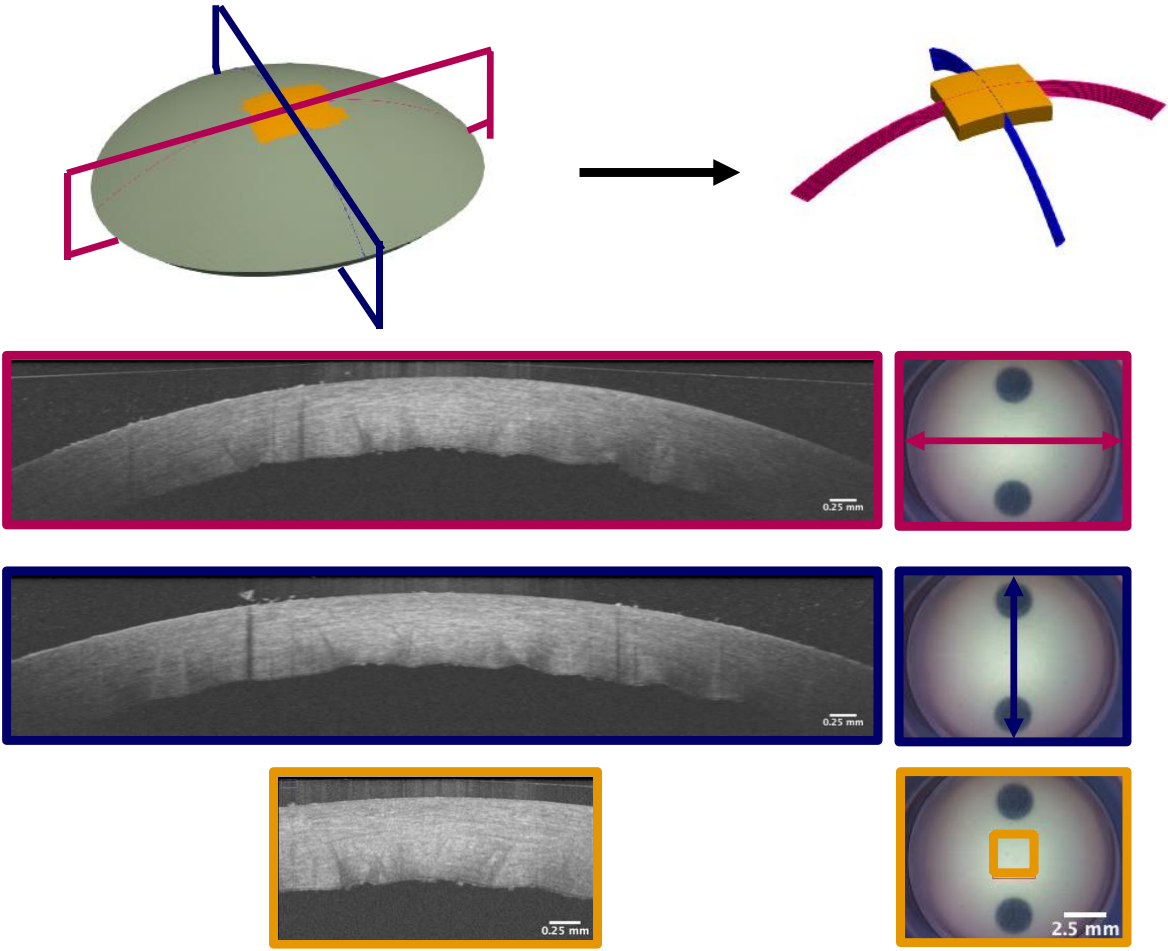

**Figure 3:**

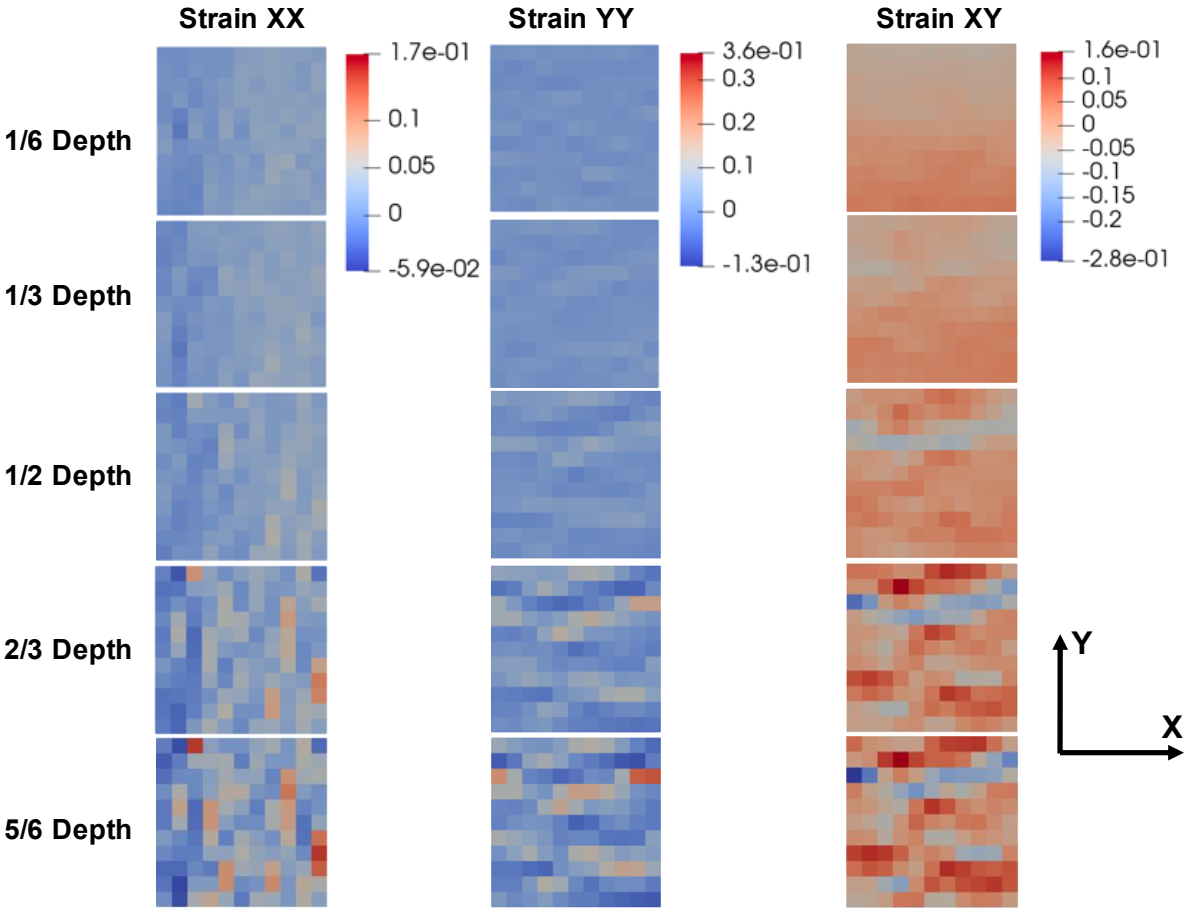

**Figure 4:**

(A)
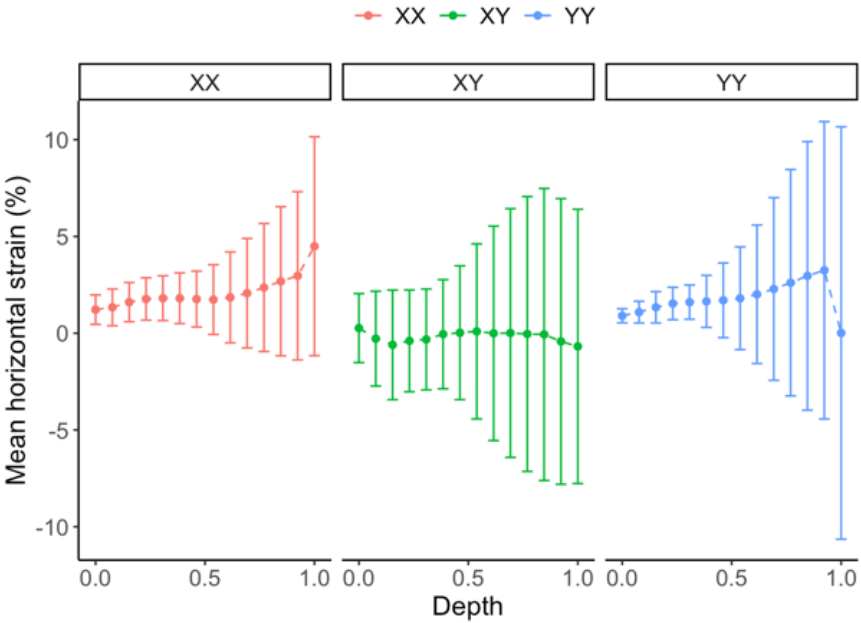

(B)
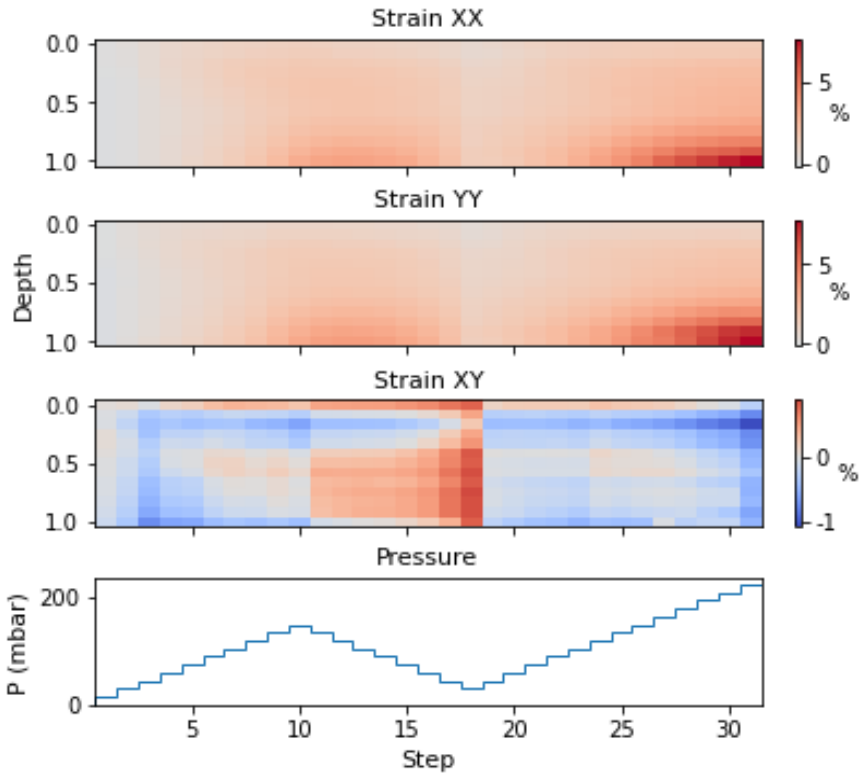

**Figure 5:**

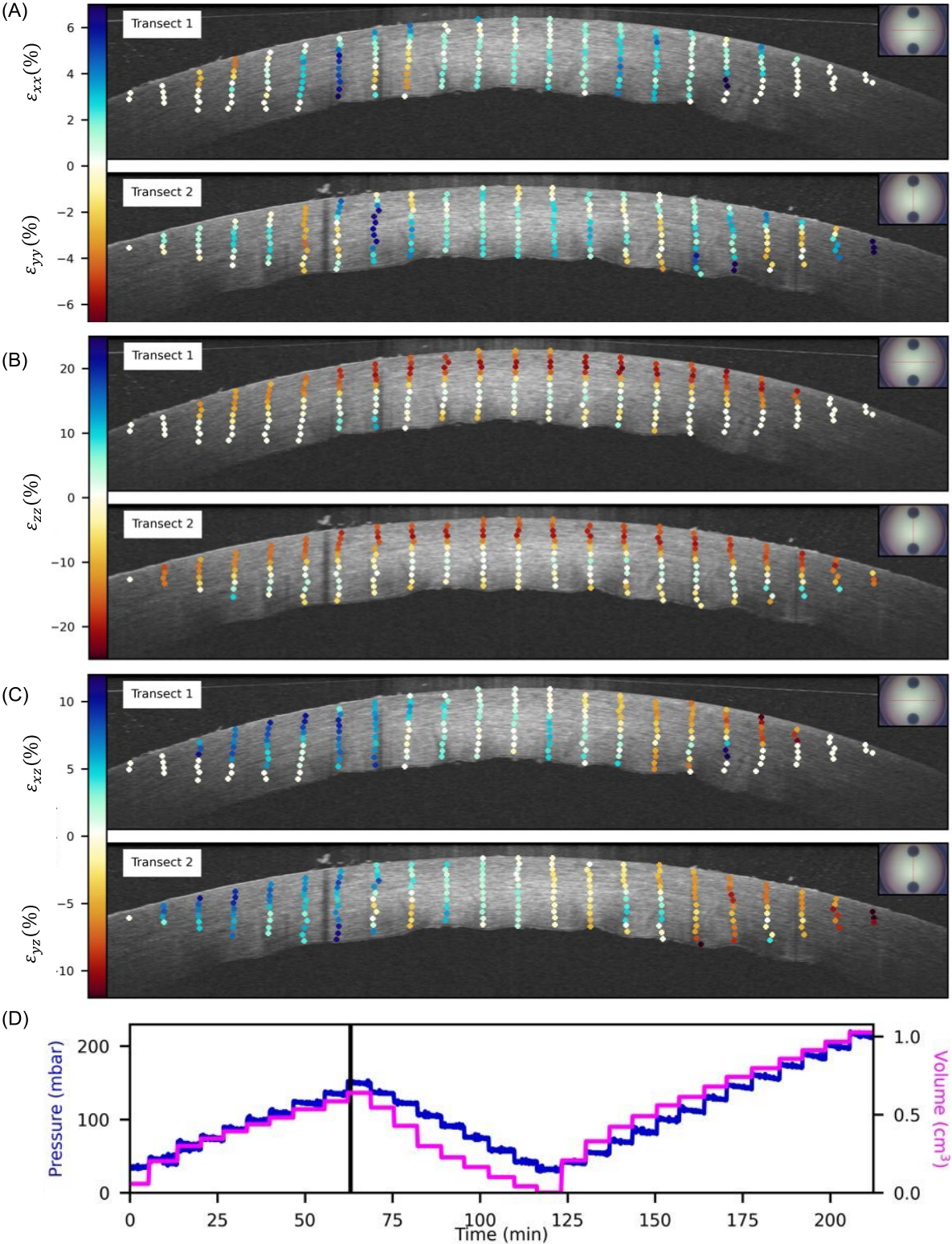

**Figure 6:**

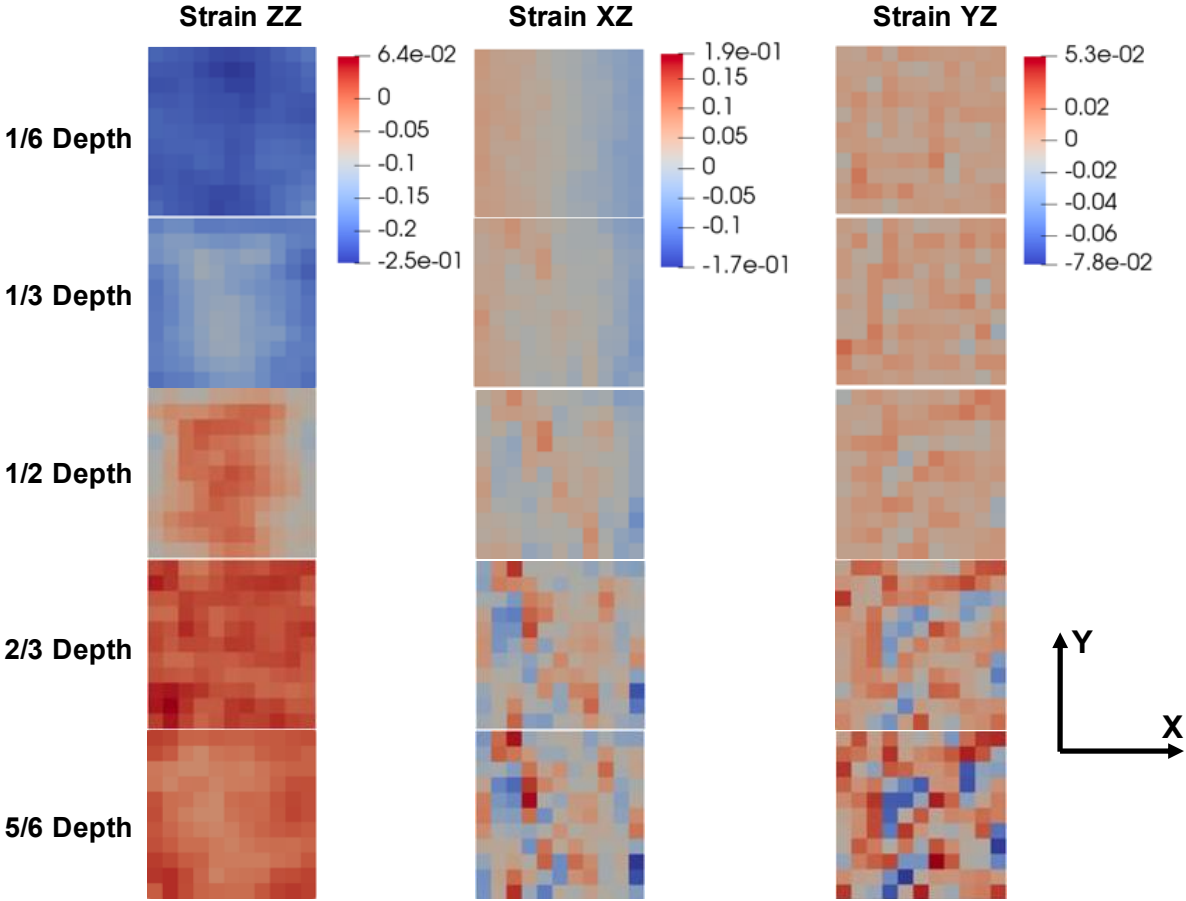

**Figure 7:**

(A)

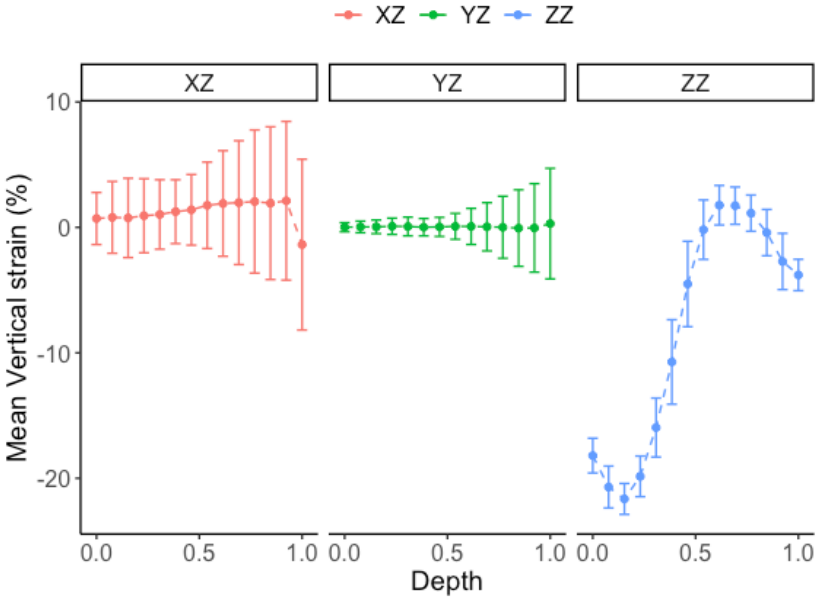

(B)

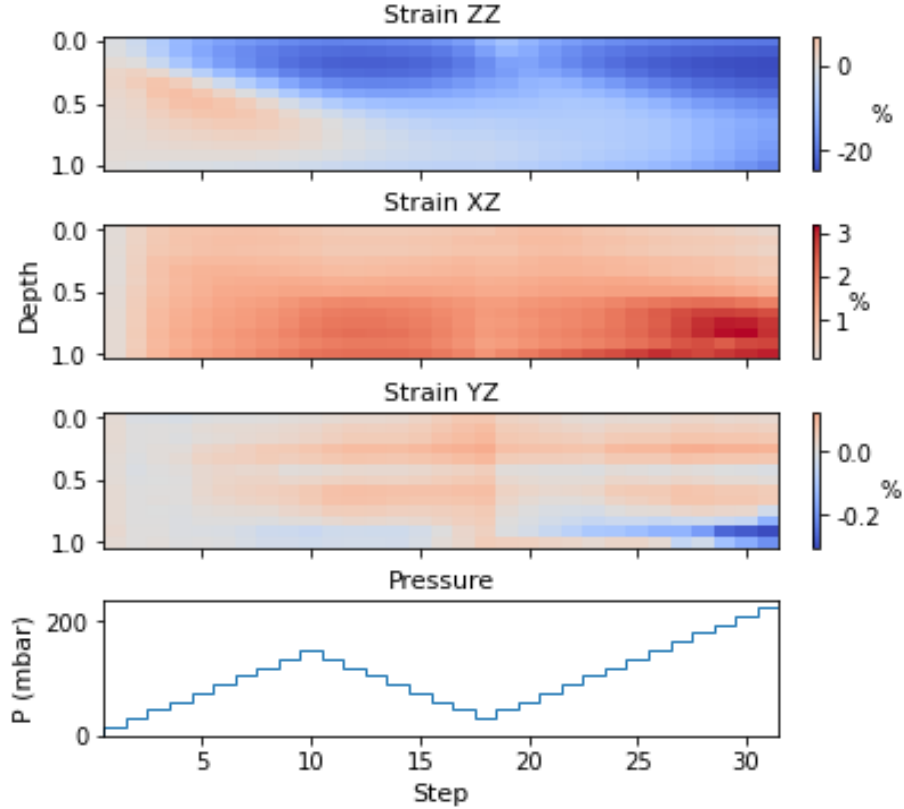